\documentclass[sigconf,nonacm,natbib=true]{acmart}
\AtBeginDocument{%
  }

\usepackage{graphicx}
\usepackage{multirow}
\usepackage{hyperref}
\usepackage{fontawesome}
\usepackage{pifont}
\usepackage{subcaption}
\usepackage{todonotes}
\usepackage{wrapfig}
\usepackage{booktabs,tabularx,array}
\usepackage{listings}
\usepackage{orcidlink}
\usepackage{xcolor}
\lstdefinestyle{promptstyle}{
  basicstyle=\ttfamily\small,
  backgroundcolor=\color{gray!10},
  frame=single,
  rulecolor=\color{gray!40},
  frameround=tttt,
  breaklines=true,
  columns=fullflexible,
  xleftmargin=1em,
  xrightmargin=1em,
  aboveskip=0.5em,
  belowskip=0.5em,
  showstringspaces=false
}

\usepackage{enumitem}
\newlist{rqlist}{itemize}{1}
\setlist[rqlist]{%
  label={},           
  leftmargin=*,       
  itemsep=1pt,
  topsep=2pt,
  parsep=0pt,
  partopsep=0pt
}

\usepackage{siunitx}
\newcolumntype{L}{>{\raggedright\arraybackslash}p{0.17\linewidth}} 
\newcolumntype{C}{>{\centering\arraybackslash}p{0.18\linewidth}}   
\newcolumntype{Y}{>{\raggedright\arraybackslash}X}     

\makeatletter
\newcommand{\acmrightssize}{\fontsize{8}{9.5}\selectfont}

\newcommand{\firstpagerights}[1]{%
  \begingroup
    \renewcommand\thefootnote{}%
    \footnotetext{%
      \acmrightssize
      \raggedright
      \setlength{\parskip}{0pt}%
      \setlength{\parindent}{0pt}%
      #1%
    }%
    \addtocounter{footnote}{0}%
  \endgroup
}
\makeatother

\begin{document}
\title[Beyond the Click: A Framework for Inferring Cognitive Traces in Search]{Beyond the Click: A Framework for Inferring \\Cognitive Traces in Search}

\author{Saber Zerhoudi}
\orcid{0000-0003-2259-0462}
\affiliation{%
  \institution{University of Passau}
  \city{Passau}
  \country{Germany}
}
\email{saber.zerhoudi@uni-passau.de}

\author{Michael Granitzer}
\orcid{0000-0003-3566-5507}
\affiliation{%
  \institution{University of Passau}
  \city{Passau}
  \country{Germany}
}
\affiliation{%
  \institution{Interdisciplinary Transformation University Austria}
  \city{Linz}
  \country{Austria}
}
\email{michael.granitzer@uni-passau.de}

\renewcommand{\shortauthors}{S. Zerhoudi et al.}

\begin{abstract}
User simulators are essential for evaluating search systems, but they primarily reproduce user actions without modeling the underlying thought process. Large-scale interaction logs record what users do, but not what they might be thinking or feeling, such as confusion or satisfaction. We present a framework for inferring cognitive traces from behavioral logs. Our method uses a multi-agent LLM system grounded in Information Foraging Theory (IFT) and validated by human experts. We annotate three public datasets (AOL, Stack Overflow, and MovieLens), producing over 530,000 cognitive labels across 50,000 sessions. A cross-dataset evaluation with a shuffled-label control reveals that cognitive labels provide the strongest signal where behavioral features are weakest: on MovieLens, the cognitive model improves F1 by up to 6.6\% over the behavioral baseline and 1.8\% above the shuffled control, while on AOL, where click patterns are highly predictive, improvements are near zero. We release the annotation collection on HuggingFace, an open-source annotation tool%
~\footnote{\label{fn:github}\url{https://github.com/searchsim-org/cognitive-traces}}~\footnote{\label{fn:website}\url{https://traces.searchsim.org/}}, and all experimental code to support future work on cognitively aware user simulation.
\end{abstract}

\ccsdesc[500]{Human-centered computing~HCI design and evaluation methods}
\ccsdesc[500]{Information systems~Users and interactive retrieval}
\ccsdesc[300]{Human-centered computing~Interactive systems and tools}

\keywords{User Simulation, Cognitive Modeling, Information Retrieval, Data Annotation, Large Language Models}

\maketitle
\firstpagerights{%
  © ACM, 2026. This is the extended author's version of the work.\\
  The definitive version was published in:
  \emph{Proceedings of the 48th European Conference on Information Retrieval (ECIR '26),
  March 29--April 2, 2026, Delft, The Netherlands}.\\
}

\begin{figure}[t]
  \centering
  \vspace{10pt}
  \includegraphics[width=1\linewidth]{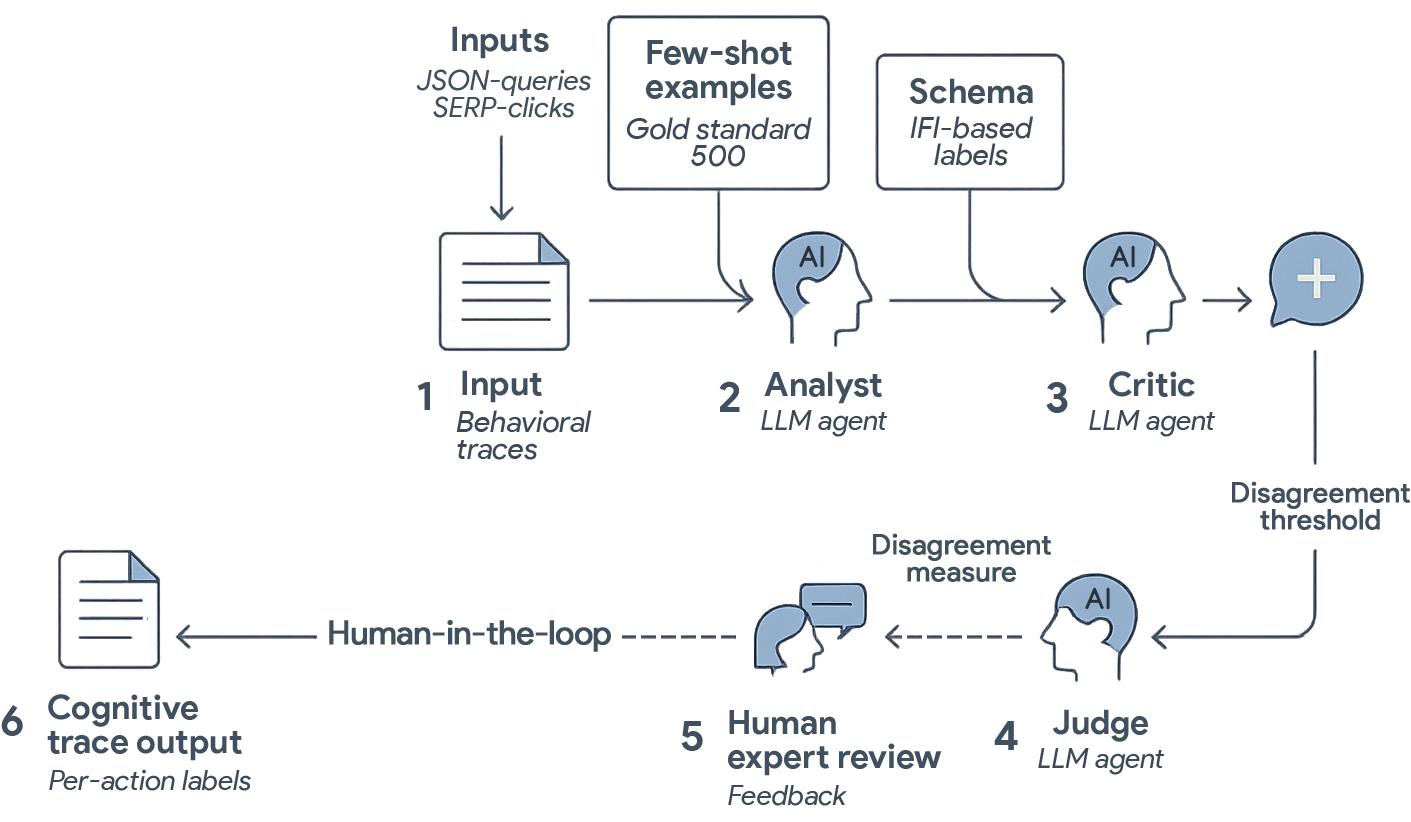}
  \caption{ The multi-agent annotation workflow. A behavioral trace is analyzed, debated, and judged to produce a final, justified cognitive trace.}
  \label{fig:annotation_workflow}
\end{figure}

\section{Introduction}

User simulation is an essential method for the evaluation of interactive information retrieval systems~\cite{Balog:2025:SIGIR,Balog:2025:ArXiv}. It offers a scalable and reproducible alternative to costly user studies, allowing researchers to test new algorithms under many conditions~\cite{Kelly:2009:FTIR}. Current simulators, often constructed using modern sequence models or reinforcement learning, have become very good at reproducing the observable actions of users~\cite{Borisov:2016:WWW,Azzopardi:2024:SIGIR-AP}. These models can generate realistic patterns of query reformulations, document selections, and other behaviors by learning from large-scale interaction logs~\cite{Jansen:2009:jasis,Chuklin:2015:synthesis}.

However, this behavioral accuracy hides a deep limitation: today's simulators are \textit{cognitively unaware}. They are effective at copying what users do but have no understanding of why they do it. The actual human process of seeking information is a complex mental activity, filled with moments of confusion, growing understanding, frustration, and eventual satisfaction~\cite{Kuhlthau:1991:jasis,Lindsay:2008:bjet,Gwizdka:2010:jasis,Pirolli:2018:db}. These internal states are not random side effects; they directly influence a user's decisions. A feeling of confusion may prompt a user to simplify their query, while a sense of making progress encourages them to continue. By ignoring this cognitive layer, our simulators remain mechanistic and incomplete, preventing us from properly assessing systems on the human-centered qualities they are meant to support.

The key to creating the more advanced user simulators is data that connects observable actions to this unobservable cognitive dimension~\cite{DiriyeWBD:2012:CIKM}. We propose that the missing component is a large-scale collection of interaction logs that is augmented with inferred traces of the user's cognitive process. 

These cognitive traces enable two major advances. First, they allow for the design of advanced simulators that learn the signs of user struggle, such as the difference between a user who is lost and one who is making progress~\cite{Sandstrom:2010:jasis,FeildAJ:2010:SIGIR,SongSWA:2014:SIGIR}. This would let us ask more specific questions about retrieval systems: Does a novel ranking algorithm reduce user frustration, even if it doesn't improve nDCG (normalized Discounted Cumulative Gain)? Does a conversational system's proactive suggestion spark curiosity or induce cognitive load? Second, they allow for new metrics like ``Session Difficulty Score'' or a ``User Frustration Rate''~\cite{Arguello:2014:ECIR,FeildAJ:2010:SIGIR}, shifting IR evaluation from a system-focused view to a user-focused one. 

In this paper, we present a concrete step toward the goal of improving simulators. We introduce a new framework and a collection of resources for inferring cognitive traces from search logs. We make the following four main contributions:

\begin{enumerate}[label=(\arabic*),labelindent=0pt,leftmargin=2em,labelsep=0.4em] 
    \item We propose a formal schema and human-in-the-loop framework for cognitive state annotation based on IFT.
    \item We release a collection of cognitive annotations for three public datasets (AOL~\cite{MacAvaney:2022:ECIR}, Stack Overflow~\cite{StackExchange:DataDump}, and MovieLens~\cite{HarperK:2016:TIIS}) on HuggingFace%
~\footnote{\label{fn:huggingface:ml}\url{https://huggingface.co/datasets/searchsim/cognitive-traces-movielens}}~\footnote{\label{fn:huggingface:aol}\url{https://huggingface.co/datasets/searchsim/cognitive-traces-aol}}~\footnote{\label{fn:huggingface:so}\url{https://huggingface.co/datasets/searchsim/cognitive-traces-stackoverflow}}, totalling over 530,000 labels across 50,000 sessions.
    \item We provide a cross-dataset experimental validation (4 tasks $\times$ 3 datasets) with a shuffled-label control, showing that cognitive labels carry genuine predictive signal, with the largest gains in domains where behavioral features are weakest.
    \item We release an open-source annotation tool and all experimental code for reproducing our results.
\end{enumerate}

\section{Related Work}

Our work draws upon several research areas: user simulation for information retrieval, the modeling of user search behavior, large-scale data annotation methodologies, and the emerging use of Large Language Models for data generation. We review each in turn to precisely situate our contribution.

\subsection{User Simulation for Information Retrieval} 

User simulators have a long history in IR, with most falling into two main groups: agenda-based and model-based~\cite{Balog:2025:ArXiv,Azzopardi:2024:SIGIR-AP}. Agenda-based simulators use a predefined set of rules and goals to direct their actions. Their main advantage is that they are easy to interpret, but they can be inflexible and have difficulty showing a wide range of user behaviors. Due to these limitations, the primary approach in recent years has been model-based simulation. These models learn user behavior directly from large interaction logs, using methods from reinforcement learning or sequence modeling to generate realistic behavioral patterns~\cite{HassanJK:2010:WSDM,Mehrotra:2017:CIKM}. Earlier work has also modeled query-change dynamics explicitly (e.g., with contextual Markov models) for simulating user search behaviour~\cite{Zerhoudi:2021:FIRE}.

While these models are effective at reproducing user actions, how they represent the user's internal state is a key weakness. A simulator's current models often represent the user through a static interest profile of their topic of interest or a simple history of their recent actions. This approach does not account for the dynamic mental shifts a person experiences during a search. Our work does not aim to replace these simulation models; instead, we aim to provide a new kind of data that can supply them with a more detailed and realistic representation of the user's internal state.

Evaluating how well simulators reproduce real interaction patterns is itself a key problem, with methods that compare behavioural properties of simulated and real sessions~\cite{Zerhoudi:2022:ECIR}.

\subsection{User Behavior Modeling and Cognitive States} 

Beyond simulation, a large amount of research in IR and Human-Computer Interaction (HCI) has focused on modeling user behavior to understand and predict their experience. Studies have shown a strong connection between implicit signals—such as clicks, query changes, and the time spent on a page—and important outcomes like user satisfaction or whether a task is completed~\cite{Hassan:2013:CIKM,Mehrotra:2017:CIKM}. These models are valuable but often treat the user's internal state as a hidden factor, not as a direct subject of study.

Most relevant to our paper is the important early research in HCI that directly measured user cognitive states during information seeking. Using detailed, small-scale lab methods like think-aloud protocols, interviews, and physiological sensors, these studies provided strong evidence of the connection between search behaviors and internal states like confusion, frustration, and cognitive load~\cite{Kuhlthau:1991:jasis,Gwizdka:2010:jasis}. This work supplied the concepts and vocabulary for understanding a user's experience, proving that the cognitive dimension is a critical factor in the interaction.

However, the methods that make these studies so insightful also make them impossible to scale to large datasets. A significant gap exists between the deep but small-scale data from HCI and the large but shallow data found in interaction logs. To our knowledge, this work is the first to directly address this gap by introducing a computational framework for inferring cognitive states at scale, combining the conceptual depth of HCI with the scale required for modern IR evaluation.

\subsection{Methods for Data Annotation} 

The creation of large datasets for IR, such as those with relevance judgments, has long depended on crowdsourcing. While this approach is effective for collecting large quantities of labels, researchers have noted the challenges involved, including maintaining data quality, addressing worker bias, and designing clear tasks~\cite{AlonsoM:2012:IPM,Lease:2013:IR}. Łajewska and Balog \cite{LajewskaBalog:CIKM2023} recently showed that collecting subjective snippet-level labels requires piloting and iterative task redesign with ongoing annotator feedback.

Our work explores a different approach. We position our language-model-based framework not as a replacement for human annotators, but as a tool to expand the reach of human expertise. We use human experts to create a high-quality initial set of examples and to verify the most difficult cases identified by our system. The language model then acts as a consistent worker that applies this expert-calibrated knowledge across millions of data points. This human-in-the-loop method seeks a balance between the scale of automation and the quality of human judgment.

\subsection{Large Language Models for Data Generation} 

The capabilities of modern language models have led to a rapid increase in their use for data generation. In IR and NLP, these models are now commonly used for tasks like creating synthetic queries to improve retrieval, generating answers for question-answering systems, and augmenting training sets~\cite{Bonifacio:2022:ArXiv,Dai:2023:ICLR}. There is also growing interest in using language models as automatic evaluators or even as complete user simulators~\cite{Zheng:2023:nips,Zhang:2024:SIGIR}.

Our work is different because we do not use the language model to create new user actions. Instead, we use it to infer a hidden cognitive layer on top of real human behavior data. This requires the model to reason about the likely motivations for a person's actions, a task that has not been explored at this scale. Our contribution is therefore not in generating new behaviors, but in adding a new dimension of understanding to existing behavioral records.

\section{Methodology}

Our main contribution is a general and reproducible framework for adding a layer of inferred cognitive traces to existing records of user behavior. The framework is designed to be grounded in theory, computationally scalable, and validated by human experts. The process has four main parts: (1) selecting and preparing a diverse set of foundational datasets; (2) using a principled schema based on Information Foraging Theory to define the cognitive labels; (3) generating the labels at scale using our multi-agent language model system; and (4) verifying and refining the output with a human-in-the-loop process.

\subsection{Foundational Datasets}
To demonstrate the flexibility of our framework, we apply it to three publicly available datasets that represent different aspects of information seeking: open-domain web search, technical question answering, and preference discovery in a recommender system. 

\begin{enumerate}[label=(\arabic*),labelindent=0pt,leftmargin=2em,labelsep=0.4em] 
    \item \textbf{AOL User Session Collection (aol-ia variant~\cite{MacAvaney:2022:ECIR}) (Web Search):} This is a large-scale log of user queries and clicks. We use the aol-ia version, which connects the original log to snapshots of the clicked web pages as they appeared around 2006, retrieved from the Internet Archive. We acknowledge the privacy issues of the original 2006 data release~\cite{Barbaro:2006:NYT}; our work uses the standard anonymized collection and only reports on aggregate findings. The key property of this dataset is that it provides the content of the documents users chose to view, but not the HTML of the search engine results page (SERP) where those links were displayed. Our framework adapts to this by making inferences based on the query sequence, the content of clicked documents, and the click choices themselves.
    \item \textbf{Stack Overflow~\cite{StackExchange:DataDump} (Technical Q\&A):} We use a public data dump of the Stack Overflow site, which contains a complete history of questions, answers, comments, and votes. This dataset is valuable since it shows the behavior of users with specific, technical information needs. The explicit feedback signals, such as upvotes and accepted answers, provide additional evidence for inferring cognitive states like problem resolution or continued confusion.
    \item \textbf{MovieLens-25M~\cite{HarperK:2016:TIIS} (Recommendations):} This dataset contains 25 million ratings and free-text tags for movies, applied by users over time. It allows us to apply our framework in a non-search context. Here, we model cognitive states related to preference formation and satisfaction. A sequence of high ratings for movies from one director can be seen as a form of successful information seeking, while a sudden low rating for an anticipated movie can signal a mismatch of expectations.
\end{enumerate}

While our initial selection of the top-used dataset in each domain was guided by the project's budget, we are committed to the long-term maintenance of the collection and plan to add annotations for other important datasets in future releases. As an immediate next step, we are applying our framework to the Archive Query Log (AQL) dataset~\cite{Reimer:2023:SIGIR}, which does contain rich SERP HTML, to create a complementary set of annotations.

\subsection{A Principled Schema for Cognitive Traces}
To avoid creating purely subjective labels, we based our annotation schema on the principles of Information Foraging Theory (IFT). IFT is a framework from cognitive science that models human information seeking using analogies from animal food foraging~\cite{Lindsay:2008:bjet}. It provides an objective, task-oriented vocabulary to describe a user's process.

In IFT, a user looks for information in information patches (like a results page or a list of movies). They are guided by information scent (cues like titles or movie genres) to find valuable items~\cite{Pirolli:2018:db}. We turned these core concepts into a set of six concrete labels for user actions. Table~\ref{tab:schema} defines each label with cross-domain examples showing how the same IFT concept applies to web search, technical Q\&A, and movie recommendations.

The IFT schema was defined by the authors based on existing literature~\cite{Sandstrom:2010:jasis}. To ensure it is consistent, we performed two pilot rounds of labeling where results were compared and the categories were fixed.

\begin{table*}[t]
  \caption{The annotation schema grounded in Information Foraging Theory, with cross-domain examples.}
  \label{tab:schema}
  \centering
  \small
  \setlength{\tabcolsep}{6pt}
  \renewcommand{\arraystretch}{1.15}
  \begin{tabularx}{\textwidth}{@{}p{0.18\textwidth} p{0.14\textwidth} p{0.12\textwidth} Y@{}}
    \toprule
    \textbf{Session Event} & \textbf{IFT Concept} & \textbf{Cognitive Label} & \textbf{Operational Definition \& Cross-Domain Examples} \\
    \midrule
    User takes a targeted, directed action
      & Following a strong scent
      & \texttt{FollowingScent}
      & The user pursues a clear information trail.
        \emph{Search:} issuing ``best espresso machine under \$500''.
        \emph{SO:} posting a focused answer to a specific question.
        \emph{ML:} rating several movies by the same director. \\ \hline
    User engages with a promising lead
      & Approaching an information source
      & \texttt{ApproachingSource}
      & The user investigates an item whose cues suggest relevance.
        \emph{Search:} clicking a result whose snippet matches the query.
        \emph{SO:} upvoting or editing an answer that addresses the problem.
        \emph{ML:} selecting a highly anticipated movie to rate. \\ \hline
    User broadens or narrows scope
      & Enriching the information diet
      & \texttt{DietEnrichment}
      & The user refines their approach by adjusting scope or strategy.
        \emph{Search:} reformulating from ``laptops'' to ``lightweight laptops for travel''.
        \emph{SO:} adding tags, commenting to clarify, or editing question scope.
        \emph{ML:} exploring a new genre after several same-genre ratings. \\ \hline
    Current results are unhelpful
      & Poor scent in the current patch
      & \texttt{PoorScent}
      & The available information does not match the user's need.
        \emph{Search:} a zero-click SERP followed by a new query.
        \emph{SO:} a question with no accepted answer despite multiple attempts.
        \emph{ML:} a low rating indicating the movie did not meet expectations. \\ \hline
    User gives up on the current path
      & Deciding to leave the patch
      & \texttt{LeavingPatch}
      & The user abandons the current direction after repeated unsuccessful attempts.
        \emph{Search:} session ends after multiple reformulations without a click.
        \emph{SO:} the user stops contributing to a thread without resolution.
        \emph{ML:} a long gap or session end after several low ratings. \\ \hline
    User finds what they needed
      & Successful foraging within the patch
      & \texttt{ForagingSuccess}
      & The user obtains the desired information or outcome.
        \emph{Search:} a direct answer on the SERP (featured snippet).
        \emph{SO:} the user's answer is accepted, or they mark a question resolved.
        \emph{ML:} a high rating (4--5 stars) signaling satisfaction. \\
    \bottomrule
  \end{tabularx}
\end{table*}

\subsection{The Multi-Agent Annotation Framework}
To generate labels at scale, we developed a multi-agent language model framework. This approach improves the quality of the output by having different agents review and challenge each other's conclusions, a form of computational quality control~\cite{Du00TM:2024:ICML}. The framework has three agents:

\begin{enumerate}[label=(\arabic*),labelindent=0pt,leftmargin=2em,labelsep=0.4em] 
  \item \textbf{The Analyst:} This agent examines the complete behavioral trace of a user session. It produces an initial set of cognitive labels for each action, along with a step-by-step justification for its choices based on the data.
  \item \textbf{The Critic:} This agent reviews the Analyst's output. Its purpose is to find inconsistencies or alternative explanations for the user's behavior. If it disagrees with the Analyst, it must propose a different label and provide its own counter-argument.
  \item \textbf{The Judge:} This agent is the final decision-maker. It considers both the Analyst's proposal and the Critic's challenge. It then makes a final choice for the label and writes a summary explanation for its decision~\cite{Zheng:2023:nips}.
\end{enumerate}


To implement this, we selected specific large language models (LLMs) for each role based on their known strengths. This model-to-role assignment was based on a preliminary study using our 500-session gold standard, where we assessed each model's qualitative strengths. We use \texttt{Claude 3.5 Sonnet}~\cite{Anthropic:2024:Claude3.5Sonnet} as the \textbf{Analyst}, as our tests confirmed its superior combination of high-speed sequential reasoning and cost-effectiveness, making it ideal for the high-volume ``first-pass'' analysis. We use \texttt{GPT-4o}~\cite{OpenAI:2024:GPT4o} as both the \textbf{Critic} and the \textbf{Judge}. Its robust general knowledge and nuanced understanding allow it to effectively challenge the Analyst’s assumptions and propose creative alternatives. For the final Judge role, this same model’s superior ability to synthesize complex arguments and weigh conflicting evidence makes it the most suitable arbiter for making a final, well-reasoned decision. The entire process, as illustrated in Figure~\ref{fig:annotation_workflow}, is guided by a structured prompt that provides each model with its persona, the task instructions, the label definitions from our schema, and the required output format.

For transparency and reproducibility, all artifacts—the datasets, annotation code, and the prompts structure—are released in our public repository\textsuperscript{\ref{fn:github}}.

For transparency and reproducibility, the core prompt structure provided to the agents is summarized in Listing~\ref{lst:system-prompt}.
\begin{lstlisting}[
  style=promptstyle,
  basicstyle=\ttfamily\footnotesize,
  caption={System prompt used in the multi-agent framework.},
  label={lst:system-prompt}
]
SYSTEM PROMPT
--------------
Persona: You are an expert in Human-Computer Interaction specializing in Information Foraging Theory.

Task: Apply a cognitive label from the provided schema to each user action in the search session.

Schema: [Detailed definitions of the 6 Information Foraging Theory labels from Table 1 are inserted here.]

Input: A JSON object containing the session's query sequence, SERP HTML for each query, and click data.

Output Format: Output a JSON list, where each item corresponds to a user action and contains two keys:
  - "label": the assigned Information Foraging Theory label
  - "justification": a 1-2 sentence explanation citing evidence 
    from the input.
\end{lstlisting}

\subsection{Human-in-the-Loop Validation}
A purely automated system is not sufficient for this complex task. We therefore integrated human expertise directly into our framework to guide and validate the process through three distinct stages.

\begin{enumerate}[label=(\arabic*),labelindent=0pt,leftmargin=2em,labelsep=0.4em] 
  \item \textbf{Gold Standard Creation:} We first created a validation set of 500 sessions. To ensure this set was diverse and included the challenging edge cases necessary for robust validation, we used a two-stage stratified sampling approach. First, we stratified the sample proportionally across our three datasets (\texttt{AOL-IA}, Stack Overflow, and MovieLens). Second, within each dataset, we further stratified by behavioral archetype, intentionally over-sampling ambiguous patterns such as ``zero-click'' sessions, ``refining'' sessions (those with many query reformulations), and ``bouncing'' sessions (those with many fast, brief clicks). Three human experts (Ph.D. students in HCI and IR) independently annotated this set, achieving a high inter-annotator agreement (Krippendorff’s $\alpha=0.78$)~\cite{hayes:2007:answering} and establishing our gold standard.
  \item \textbf{Framework Calibration and Validation:} The 500-session gold standard serves two distinct purposes. First, we manually selected 3-5 archetypal sessions from this set—sessions that are exceptionally clear examples of each cognitive state—to include as fixed few-shot examples in our agents' prompts. This provides clear, high-quality demonstrations of the task. Second, the entire 500-session set was used as a held-out testbed to validate our complete framework. We tuned the prompts and agent interactions until our framework achieved a 92.4\% accuracy against the human-generated gold labels, confirming its high fidelity before the large-scale annotation run.
  \item \textbf{Active Learning for Quality Assurance:} During the large-scale annotation run, our system automatically flags the 1\% of cases where the Analyst and Critic have the strongest disagreement (measured by semantic distance between their justifications). These most difficult cases are then sent to our human experts for a final ruling. This active learning loop ensures that human attention is focused where it is most needed, improving the overall quality of the final dataset~\cite{Xu:2013:icimcs,Seung:1992:colt}.
\end{enumerate}

\section{Experimental Validation}

We evaluate whether inferred cognitive traces provide genuine predictive utility beyond what behavioral features alone can achieve. To test this rigorously, we design a cross-dataset evaluation spanning three domains with fundamentally different interaction vocabularies. Every experiment includes a \textbf{shuffled-label control}: a model with the same architecture as the cognitive model, but whose IFT labels are randomly permuted within each session. If the cognitive model outperforms both the behavioral baseline \emph{and} the shuffled control, the improvement comes from the specific content of the labels rather than from an additional embedding dimension.

\subsection{Task Design}

We define four binary prediction tasks that apply across all three datasets. Each task uses purely behavioral targets. No task defines its target from cognitive labels, which ensures the cognitive model cannot trivially reconstruct the answer.

\paragraph{Session Continuation.}
Given a session prefix up to event $t$ ($t \geq 2$), predict whether the session continues ($y{=}1$) or ends ($y{=}0$). This purely positional target tests whether cognitive state trajectories predict user disengagement.

\paragraph{Positive Outcome.}
Predict whether the user's next interaction is positive. The definition of ``positive'' varies by domain: in AOL, a click (vs.\ a query reformulation); in Stack Overflow, posting an answer (vs.\ a comment); in MovieLens, a rating ${\geq}\,4.0$ (vs.\ ${\leq}\,2.5$).

\paragraph{Exploration vs.\ Exploitation.}
Predict whether the user's next action explores new territory or deepens within the current path. This task connects directly to the IFT concepts of patch leaving and diet enrichment (exploration) versus scent following and source approaching (exploitation). AOL uses query Jaccard similarity, Stack Overflow uses thread continuity, and MovieLens uses genre overlap between consecutive ratings.

\paragraph{Session Success from Early Context.}
Given only the first three events, predict full-session success. AOL defines success as completing the search in three or fewer queries. Stack Overflow requires the session to contain a posted answer. MovieLens requires any rating of 4.0 or higher. Sessions must have at least six events.

\begin{figure}[t]
\centering
\includegraphics[width=\linewidth]{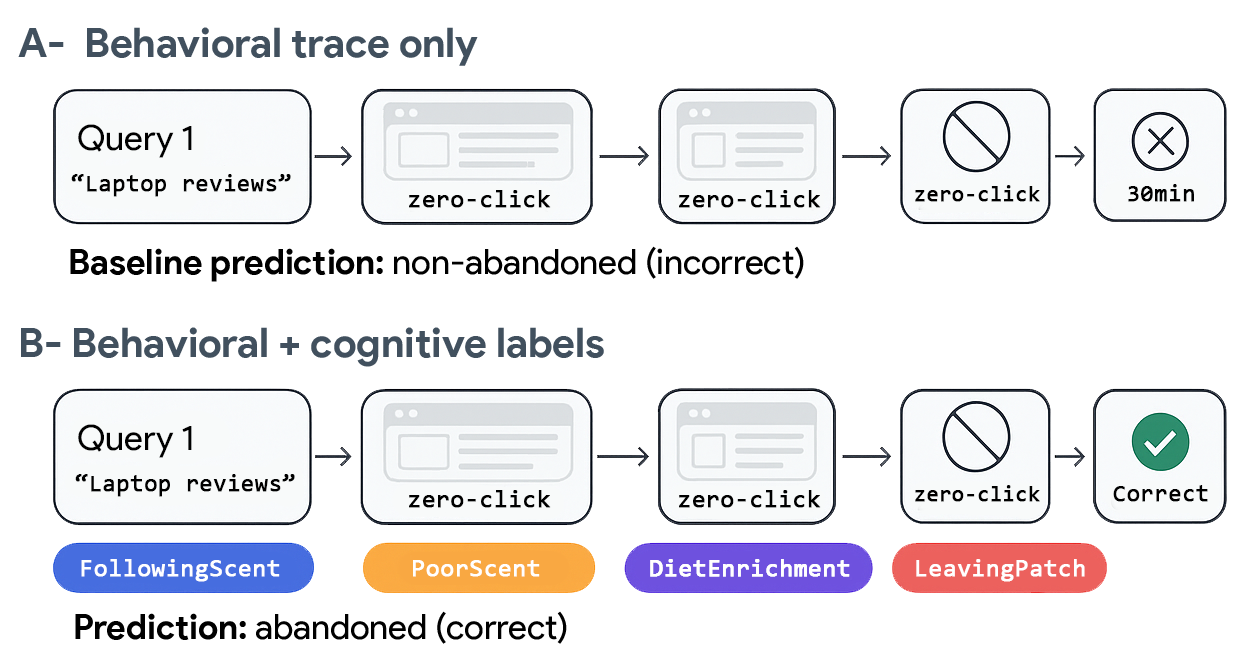}
\caption{An example session from AOL. (A) The behavioral trace shows queries and zero-click events but predicts non-abandonment. (B) Cognitive labels reveal a trajectory from \texttt{\small FollowingScent} through \texttt{\small PoorScent} to \texttt{\small LeavingPatch}, correctly indicating failure.}
\label{fig:session-example}
\end{figure}

\subsection{Experimental Setup}

\paragraph{Model Architecture.}
All three model variants share a 4-layer Transformer encoder (768 hidden units, 8 attention heads, 3072 FFN) with CLS-token pooling and a linear classification head. Text is encoded by S-BERT (\texttt{all-MiniLM-L6-v2}, 384 dimensions). Action-type embeddings (32 dimensions) are learned per dataset, with vocabulary sizes that vary by domain (AOL: 4, Stack Overflow: 12, MovieLens: 4, each including a PAD token). The \textbf{Behavioral Baseline} uses a 416-dimensional input per event (384 text + 32 action type). The \textbf{Cognitive-Enhanced Model} adds a 32-dimensional IFT label embedding (7 classes: 6 IFT labels + PAD), for a 448-dimensional input. The \textbf{Shuffled-Label Control} uses the same architecture, but randomly permutes label IDs within each non-padded sequence during the forward pass.

\paragraph{Training.}
Adam optimizer, learning rate $5 \times 10^{-5}$, batch size 64, maximum 20 epochs with early stopping (patience 7 on validation F1). Binary cross-entropy loss with class-balanced training sets via majority-class downsampling. User-based splits (80/10/10) prevent information leakage across splits. All experiments run on a single NVIDIA A100 GPU with seed 42.

\subsection{Results}

Table~\ref{tab:cross-dataset} presents the full cross-dataset evaluation across 12 experiments. We report F1 and AUC for each model, along with two improvement measures: $\Delta_{\text{B}}$, the cognitive model's improvement over the behavioral baseline, and $\Delta_{\text{S}}$, the improvement over the shuffled control. A positive $\Delta_{\text{S}}$ indicates genuine cognitive signal.

\begin{table*}[t]
  \caption{Cross-dataset evaluation: F1 and AUC on held-out test sets for three model variants. \textbf{Bold} marks the best F1 per row. $\Delta_{\text{B}}$ = cognitive improvement over behavioral baseline; $\Delta_{\text{S}}$ = cognitive improvement over shuffled control. MovieLens consistently shows the strongest genuine cognitive gains, while AOL operates at or near ceiling.}
  \label{tab:cross-dataset}
  \centering
  \small
  \setlength{\tabcolsep}{4pt}
  \begin{tabular}{ll ccc cc cc}
    \toprule
    & & \multicolumn{3}{c}{\textbf{F1 Score}} & \multicolumn{2}{c}{\textbf{F1 Change}} & \multicolumn{2}{c}{\textbf{AUC}} \\
    \cmidrule(lr){3-5} \cmidrule(lr){6-7} \cmidrule(lr){8-9}
    \textbf{Task} & \textbf{Dataset}
      & {Behav.} & {Cogn.} & {Shuf.}
      & {$\Delta_{\text{B}}$} & {$\Delta_{\text{S}}$}
      & {Behav.} & {Cogn.} \\
    \midrule
    \multirow{3}{*}{Sess.\ Contin.}
      & AOL & 0.953 & 0.956 & 0.955 & +0.3\% & +0.1\% & 0.967 & 0.991 \\
      & SO  & 0.660 & \textbf{0.669} & 0.667 & +1.4\% & +0.4\% & 0.762 & 0.768 \\
      & ML  & 0.655 & \textbf{0.698} & 0.695 & +6.6\% & +0.4\% & 0.728 & 0.771 \\
    \midrule
    \multirow{3}{*}{Pos.\ Outcome}
      & AOL & 0.955 & 0.958 & 0.957 & +0.3\% & +0.1\% & 0.982 & 0.983 \\
      & SO  & 0.663 & 0.655 & \textbf{0.670} & $-$1.1\% & $-$2.2\% & 0.625 & 0.627 \\
      & ML  & 0.667 & \textbf{0.709} & 0.703 & +6.3\% & +0.8\% & 0.481 & 0.744 \\
    \midrule
    \multirow{3}{*}{Expl./Exploit.}
      & AOL & 0.719 & 0.738 & \textbf{0.741} & +2.7\% & $-$0.5\% & 0.778 & 0.812 \\
      & SO  & 0.711 & \textbf{0.722} & 0.718 & +1.5\% & +0.5\% & 0.797 & 0.797 \\
      & ML  & 0.667 & \textbf{0.679} & 0.667 & +1.8\% & +1.8\% & 0.552 & 0.686 \\
    \midrule
    \multirow{3}{*}{Early Success}
      & AOL & 0.667 & 0.667 & 0.667 & 0.0\% & 0.0\% & 0.501 & 0.542 \\
      & SO  & \textbf{0.919} & 0.919 & 0.919 & 0.0\% & 0.0\% & 0.938 & 0.946 \\
      & ML  & 0.667 & 0.773 & \textbf{0.785} & +15.9\% & $-$1.5\% & 0.598 & 0.873 \\
    \bottomrule
  \end{tabular}
\end{table*}

Three findings emerge from these results.

\paragraph{Finding 1: Cognitive labels help most where behavioral signals are weakest.}
MovieLens shows the strongest cognitive improvements across all four tasks, with F1 gains of +1.8\% to +6.6\% over the behavioral baseline. On three of four tasks, the cognitive model also exceeds the shuffled control (+0.4\% to +1.8\%), confirming genuine cognitive signal. Table~\ref{tab:action-types} explains why: MovieLens has only three action types (\texttt{RATE}, \texttt{BELIEF\_ELICIT}, \texttt{BELIEF\_PREDICT}), none of which encode whether a user enjoyed a movie. A behavioral model simply cannot tell high ratings from low ratings using action types alone. Cognitive labels fill this gap because IFT states like \texttt{ForagingSuccess} correlate with satisfaction, while \texttt{PoorScent} correlates with dissatisfaction.

AOL also has only three action types, but the query, click, and SERP view cycle is inherently predictive of user intent. The behavioral model reaches F1 above 0.95 on two tasks, leaving almost no room for improvement. This shows that the raw count of action types does not determine cognitive utility. What matters is whether the available actions are informative enough for the prediction target.

\begin{table}[t]
  \caption{Cross-dataset summary. MovieLens has few action types and none that encode content quality, making cognitive labels essential. AOL's click patterns are already highly informative, so cognitive labels add little.}
  \label{tab:action-types}
  \centering
  \small
  \setlength{\tabcolsep}{4pt}
  \begin{tabular}{lccc}
    \toprule
    \textbf{Dataset} & \textbf{\# Action Types} & \textbf{Avg $\Delta_\text{B}$ F1} & \textbf{Avg $\Delta_\text{S}$ F1} \\
    \midrule
    MovieLens & 3 & \textbf{+7.7\%} & \textbf{+0.4\%} \\
    Stack Overflow & 11 & +0.5\% & $-$0.4\% \\
    AOL & 3 & +0.8\% & $-$0.1\% \\
    \bottomrule
  \end{tabular}
\end{table}

\paragraph{Finding 2: The shuffled-label control is essential for evaluation.}
On several tasks, the shuffled model matches or exceeds the cognitive model. For AOL Explore/Exploit, the shuffled model reaches F1 of 0.741 while the cognitive model reaches 0.738. For MovieLens Early Success, shuffled F1 is 0.785 versus cognitive F1 of 0.773. Without the shuffled control, these gains would be falsely attributed to cognitive content. The ablation shows that some improvements come from the regularization effect of adding any additional embedding, regardless of its content. We consider only tasks where cognitive exceeds shuffled as evidence of genuine cognitive signal.

\paragraph{Finding 3: AUC improvements exceed F1 improvements.}
The largest improvements appear in AUC rather than F1. MovieLens Positive Outcome shows a +54.7\% AUC improvement (0.481 to 0.744) while F1 improves by +6.3\%. This pattern suggests that cognitive labels improve probability calibration and ranking quality more than classification accuracy at a fixed threshold. For applications where calibrated probabilities matter, such as ranking or intervention systems, cognitive traces may provide greater practical value than the F1 numbers alone suggest.

\subsection{Discussion}

\paragraph{When Do Cognitive Labels Help?}
Our evaluation reveals a clear pattern: cognitive labels provide the most value when the behavioral action vocabulary is informationally insufficient for the prediction target. MovieLens's three action types encode what a user did (rated a movie, answered a belief question) but not how they felt about the content. IFT labels supply this missing signal by encoding satisfaction (\texttt{ForagingSuccess}), engagement (\texttt{ApproachingSource}), and frustration (\texttt{PoorScent}). In contrast, AOL's query, click, and SERP view cycle is already so predictive of user intent that cognitive labels add negligible information.

This finding has practical implications. Researchers should prioritize cognitive annotation for datasets and tasks where behavioral signals are weak or ambiguous. In domains with rich implicit feedback (such as dwell time or click-through rates), cognitive traces may be redundant. In domains with sparse or uninformative actions (such as ratings or yes/no responses), they may be essential. This pattern was not visible in earlier single-dataset evaluations that lacked a cross-domain comparison.

\paragraph{Assessment of Effect Sizes.}
We report both $\Delta_\text{B}$ (improvement over behavioral baseline) and $\Delta_\text{S}$ (improvement over shuffled control). The latter is the more rigorous measure. Across the viable cross-dataset experiments (excluding degenerate cases where all models predict the majority class), the average $\Delta_\text{S}$ is modest. MovieLens accounts for nearly all genuine signal. We urge caution when interpreting $\Delta_\text{B}$ alone, since it conflates genuine cognitive content with the regularization effect of adding any embedding.

\paragraph{Relationship to Preliminary Results.}
A preliminary version of this work evaluated cognitive traces on two AOL-only tasks (session outcome forecasting and struggle recovery prediction) using a text-only behavioral baseline without action-type embeddings. That evaluation reported larger improvements because the baseline could not leverage the inherent structure of the query--click--SERP cycle. The present evaluation strengthens all baselines with action-type embeddings and introduces a shuffled-label control, producing a fairer comparison. Under these conditions, AOL improvements are moderate (consistent with our finding that click patterns are already highly predictive in web search) while the largest genuine gains appear on MovieLens, where behavioral features are least informative.

\paragraph{Predictive Utility, Not Causality.}
These experiments show that cognitive traces carry predictive signal for behaviorally defined outcomes. They do not establish that cognitive states cause behavioral changes. Establishing causality would require intervention studies or causal annotation protocols where each event is labeled using only past context.

\section{Annotation Collection and Tools}

To ensure our work provides a useful and lasting contribution to the community, we release all artifacts as open-source resources. These include our collection of cognitive annotations for the three benchmark datasets, the source code for our annotation framework, and a general-purpose software tool for annotating new datasets.

\subsection{Resource Overview and Access}

The complete set of resources is publicly available\textsuperscript{\ref{fn:github}} and contains four main components:

\begin{enumerate}[label=(\arabic*),labelindent=0pt,leftmargin=2em,labelsep=0.4em] 
    \item \textbf{The Cognitive Annotation Collection:} A set of files containing the inferred cognitive labels for all user sessions we processed from the AOL-IA, Stack Overflow, and MovieLens datasets.
    \item \textbf{The Annotator Tool:} A web-based, interactive tool designed to help researchers apply our cognitive annotation schema (or their own) to any session-based dataset.
    \item \textbf{The Pre-Trained Model:} A lightweight, fine-tuned model that predicts cognitive labels. This model is integrated into the Annotator tool but is also available as a standalone file for independent use.
    \item \textbf{Source Code and Documentation:} The full source code for our multi-agent framework (for reproducibility) and the Annotator tool, along with detailed documentation and usage examples.
\end{enumerate}

\subsection{The Cognitive Annotation Collection}

Our initial release provides cognitive annotations for three large-scale and diverse datasets, demonstrating the flexibility of our framework across different information-seeking domains. Table~\ref{tab:resources} summarizes the scale of the annotations.

\begin{table}[t]
  \caption{Cognitive annotation collection statistics. All three datasets are released on HuggingFace for direct use with standard ML pipelines.}
  \label{tab:resources}
  \centering
  \small
  \setlength{\tabcolsep}{4pt}
  \begin{tabular}{lcccc}
    \toprule
    \textbf{Dataset} & \textbf{Domain} & \textbf{Sessions} & \textbf{Events} & \textbf{Actions} \\
    \midrule
    AOL-IA & Web Search & 22,039 & 245,786 & 3 \\
    Stack Overflow & Technical Q\&A & 18,629 & 175,326 & 11 \\
    MovieLens & Recommender & 10,274 & 111,561 & 3 \\
    \midrule
    \textbf{Total} & & \textbf{50,942} & \textbf{532,673} & \\
    \bottomrule
  \end{tabular}
\end{table}

\begin{figure}[t]
\centering
\includegraphics[width=\linewidth]{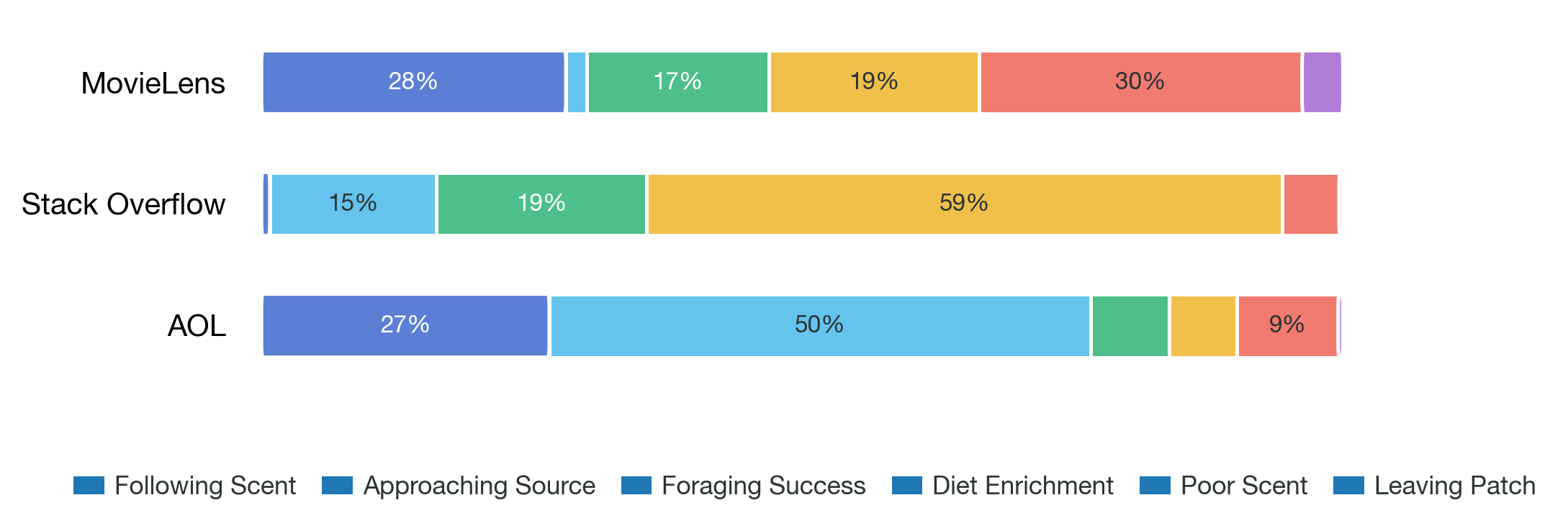}
\caption{Distribution of inferred cognitive states across the three annotated datasets. The dominant label varies by domain: clicks drive \texttt{ApproachingSource} in AOL, iterative editing drives \texttt{DietEnrichment} in Stack Overflow, and mixed satisfaction drives a balanced distribution in MovieLens.}
\label{fig:label-dist}
\end{figure}

The distribution of inferred cognitive states varies meaningfully across datasets (Figure~\ref{fig:label-dist}). In AOL, the most frequent labels are \texttt{ApproachingSource} (50\%) and \texttt{FollowingScent} (27\%), reflecting the click-heavy nature of web search. Stack Overflow shows a high concentration of \texttt{DietEnrichment} (59\%), consistent with iterative knowledge building through comments and edits. MovieLens exhibits a more balanced distribution, with \texttt{PoorScent} (30\%) and \texttt{FollowingScent} (28\%) as the leading labels, reflecting the exploratory nature of movie preference formation.

The annotations are provided as CSV files (and as Parquet on HuggingFace)\textsuperscript{ \ref{fn:huggingface:ml} \ref{fn:huggingface:aol} \ref{fn:huggingface:so}}, where each row represents a single user event and includes the following key fields:
\begin{itemize}[label=\textendash,labelindent=0pt,leftmargin=2em,labelsep=0.4em]
    \item \texttt{session\_id}: Unique identifier for the session.
    \item \texttt{event\_timestamp}: Timestamp of the event.
    \item \texttt{action\_type}: Type of action (e.g., \texttt{QUERY}, \texttt{CLICK}, \texttt{RATE}).
    \item \texttt{content}: Text content of the event (query text, post body, or movie/rating data).
    \item \texttt{cognitive\_label}: Final IFT label assigned by the Judge agent.
    \item \texttt{\small analyst\_justification}, \texttt{\small critic\_justification}, \texttt{\small judge\_justi-\\fication}: Full reasoning chain from all three agents.
    \item \texttt{confidence\_score}: Framework confidence score (0 to 1).
\end{itemize}

All three datasets can be loaded directly via the HuggingFace \texttt{datasets} library:

\begin{lstlisting}[style=promptstyle,basicstyle=\ttfamily\footnotesize]
from datasets import load_dataset
ds = load_dataset("searchsim/cognitive-traces-movielens")
\end{lstlisting}

\subsection{The \textit{Annotator} Tool}

A key contribution of this work is not just a static dataset but a reusable tool that enables other researchers to create their own cognitively annotated logs. We release the \textit{Annotator}, a web-based dashboard for cognitive annotation that transforms our method from a complex data pipeline into an accessible, interactive process.


The tool is designed for ease of use and offers several core features:
\begin{itemize}[label=\textendash,labelindent=0pt,leftmargin=2em,labelsep=0.4em]
    \item \textbf{Flexible Data Ingestion:} Users can upload session-based logs in CSV or JSON format and map their data columns to required fields (session ID, timestamp, query text, etc.).
    \item \textbf{AI-Assisted Annotation:} The tool includes a lightweight pre-trained model that automatically provides a suggested cognitive label and confidence score for each event.
    \item \textbf{Human-in-the-Loop Workflow:} Annotators review, accept, or correct AI suggestions, dramatically speeding up the annotation process compared to fully manual labeling.
    \item \textbf{Interactive Visualization:} Each session is displayed as a visual timeline, helping annotators see context across queries and clicks.
    \item \textbf{Easy Export:} Completed annotations can be exported in a clean, analysis-ready format with one click.
\end{itemize}

This tool effectively lowers the barrier to cognitive user-behavior research, enabling both individuals and small research groups to apply cognitive annotation to new or private datasets.

\section{Discussion and Future Work}
Our work introduces a framework for applying a cognitive layer to behavioral logs. In this section, we discuss the limitations of this approach and outline the new research directions it makes possible.

\subsection{Limitations}


\paragraph{Inferences, Not Measurements.}
The cognitive traces are plausible, theory-grounded inferences rather than direct measurements of mental states. A label like \texttt{PoorScent} is a strong indicator of difficulty, but it remains an interpretation. The ground truth of a user's thought process is inaccessible without physiological measurement or think-aloud protocols.


\paragraph{Data Richness.}
The quality of cognitive inferences depends on the richness of the input data. Our work with AOL demonstrates that the framework can operate without access to SERP content, but richer contextual data would provide stronger evidence for inferring states like \texttt{PoorScent} or \texttt{ForagingSuccess}. As a next step, we are applying the framework to the Archive Query Log (AQL) dataset~\cite{Reimer:2023:SIGIR}, which contains full SERP HTML.

\paragraph{Domain Coverage.}
We validate across web search, technical Q\&A, and movie recommendations. These three domains do not cover all information-seeking activities. The specific cognitive patterns may differ in legal research, medical information seeking, or complex creative tasks.

\subsection{Future Research Directions}

We believe this work opens up several new avenues for research.

\paragraph{1. Richer Cognitive Representations:} Our current work uses a discrete set of six labels. A natural next step is to explore more complex representations. The natural language justifications produced by our ``Judge'' agent could be used to train models that output a continuous vector in a learned ``cognitive space''. This would allow for a more nuanced understanding of user states, capturing the subtle differences between mild uncertainty and deep confusion.

\paragraph{2. Longitudinal User Modeling:} Our annotation collection, particularly for datasets like MovieLens and Stack Overflow, provides data on the same users over long periods. This allows for the study of how a user's cognitive patterns and information-seeking strategies change over time. We can now ask questions like: Do users become more efficient foragers in a domain as they gain expertise? How do their cognitive patterns shift when their interests change?

\paragraph{3. Building Cognitively-Aware Systems:} The ultimate goal of this research is to create better information access systems. Our Annotator tool and its lightweight model could be integrated into a live system to infer a user's cognitive state in real-time. This would permit the creation of systems that can proactively intervene. For example, a system that detects a user is in a repeated \texttt{PoorScent} state could offer to reformulate their query or suggest a different search strategy, helping them before they reach the point of abandonment. 

\section{Conclusion}

In this paper, we addressed a key limitation in user modeling: the gap between observable user behavior and the unobservable cognitive states that guide it. We presented a complete framework for inferring these states from behavioral logs, grounding our method in Information Foraging Theory and validating it with human experts.

Our cross-dataset evaluation across 12 experiments and three domains reveals a clear pattern: cognitive traces provide the strongest predictive signal where behavioral features are least informative. On MovieLens, where action types carry no information about content quality, cognitive labels improve F1 by up to 6.6\% with genuine signal above a shuffled-label control. On AOL, where click patterns are highly predictive, the behavioral baseline is already near ceiling. This finding provides a practical guide for the community: cognitive annotation is most valuable in behaviorally sparse domains.

By releasing cognitive annotations for three major datasets on HuggingFace, together with our open-source annotation tool and experimental code, we provide the community with both data and instruments. We hope these resources support a new generation of user simulators and information systems that account for the user's internal experience.


%
%
\bibliographystyle{ACM-Reference-Format}
\bibliography{sources}

\end{document}